\documentstyle[12pt]{article}
\newcommand{\sect}[1]{\section{#1}\setcounter{equation}{0}}

\newcommand{\eq}{\begin{equation}}
\newcommand{\eqe}{\end{equation}}
\newcommand{\dis}{\begin{displaymath}}
\newcommand{\dise}{\end{displaymath}}
\newcommand{\eqa}{\begin{eqnarray}}
\newcommand{\eqae}{\end{eqnarray}}
\newcommand{\e}[1]{\label{eq:#1}}
\newcommand{\ee}[1]{(\ref{eq:#1})}

\newcommand{\apm}{\alpha'}

\begin{document}\bigskip
\hskip 3.7in\vbox{\baselineskip12pt
\hbox{NSF-ITP-96-144}\hbox{hep-th/9612146}}
\bigskip\bigskip\bigskip

\centerline{\large \bf A Correspondence Principle for }
\centerline{\large \bf Black Holes and Strings }

\bigskip\bigskip

\centerline{\bf Gary T. Horowitz} 
\medskip
\centerline{Physics Department}
\centerline{University of California}
\centerline{Santa Barbara, CA. 93106}
\medskip
\centerline{email: gary@cosmic.physics.ucsb.edu}
\bigskip
\centerline{\bf Joseph Polchinski}
\medskip
\centerline{Institute for Theoretical Physics}
\centerline{University of California}
\centerline{Santa Barbara, CA. 93106}
\medskip
\centerline{e-mail: joep@itp.ucsb.edu}
\bigskip\bigskip

\begin{abstract}
\baselineskip=16pt

For most black holes in string theory, the Schwarzschild radius
in string units decreases as the string coupling is reduced.  We
formulate a correspondence principle, which states that (i) when the size
of the horizon drops below the size of a string, the typical black hole
state becomes a typical state of strings and D-branes with the same
charges, and (ii) the mass does not change
abruptly during the transition.  This provides a statistical interpretation
of black hole entropy.
This approach does not yield the numerical coefficient, but
gives the correct dependence on mass and charge in a wide range of cases,
including neutral black holes.
\end{abstract}
\newpage
\baselineskip=18pt

\sect{Introduction}

A few years ago, Susskind proposed \cite{suss1} that there was a
one-to-one correspondence between  Schwarzschild
black holes and fundamental string
states.  This is based on the fact that as one increases the string
coupling, the size of a highly excited string state becomes less than its
Schwarzschild radius, so it must become a black hole.  Conversely, as one
decreases the coupling, the size of a black hole eventually becomes less
than the string scale.  At this point, the metric is no longer well defined
near the horizon, so it can no longer be interpreted as a black hole.
Susskind suggested that the configuration should be described in terms of
some string state.  At large values of the mass, the typical state
consists of a small number of highly excited strings, so the black hole
should reduce to such a state at weak coupling.  In fact, the
single-string entropy approximates the total entropy up to subleading
terms, and so one can focus on states of a single highly-excited string. 
Further evidence for this correspondence between
black holes and excited string states has recently been given
\cite{hrs,suss2}.

It is widely believed that there is a discrepancy between the entropy of
a free fundamental string (which is proportional to the mass of the string
state) and the Bekenstein-Hawking entropy (which is proportional to the
{\it square} of the mass of the black hole). This apparent discrepancy
must clearly be resolved in order for the proposed correspondence to be
valid. Susskind has suggested that a large gravitational redshift might
account for the difference.

We will show that the standard formulas for the string and black hole
entropies can be related directly to one another.  More generally,
when the black hole carries Ramond-Ramond charge, the weak coupling
limit involves D-branes \cite{pol}. The correspondence between black holes at
strong coupling and
strings and D-branes at weak coupling can be stated in terms of the following
principle:

$(i)$ When the curvature at the horizon of a black hole (in the string
metric) becomes greater than
the string scale,  the typical black hole state becomes a typical state
of strings and D-branes with the same charges and angular momentum.

$(ii)$ The mass changes by at most a factor of order unity during the
transition.

The condition on the curvature in $(i)$ is just the condition for 
$\apm$ corrections to become important near the horizon, so this is a natural 
point for the transition to occur.
An immediate consequence of this  principle is that the black hole entropy  
must be comparable to the string entropy. One may wonder how this is
possible, since we explicitly assume that
the mass does not  change significantly during the transition to a string
state.
 The point is that as the
string coupling
$g$ is varied, the mass of a black hole is constant in Planck units, while
the mass of a string (ignoring gravitational corrections) is constant in
string units. So they can agree for only one value of $g$. It is natural to
equate the masses at the value of $g$ when the black
hole becomes a string.  By the above principle, this occurs 
when the size of the horizon is of order the string scale.
We will show that when the black
hole mass and string mass are set equal at this scale, their respective
entropies are also equal, up to a factor of order unity that depends on
exactly when the black hole forms.  Turning the argument around,
if we follow a given
state (that is, fixing the entropy\footnote{For a system with rapidly growing
density of states, a narrow band of states can be labeled by its entropy.}) 
adiabatically through the transition
between a black hole and a string, its mass changes by a factor which is
of order one, rather than being parametrically large.

For the Schwarzschild black hole in four dimensions, the equality of the
single-string and black hole entropies for string-sized black holes was pointed
out by Susskind~\cite{suss1}.  We extend this to all dimensions and to black
holes carrying a variety of charges. (Adding angular momentum typically
changes the entropy by at most a factor of two, so its effect is difficult to
see just from the correspondence principle.) In some cases the typical state on
the weak coupling side is a single long string, but in others it is gas
of massless strings on D-branes; two distinctly different kinds of gas
(free and interacting) arise.  Thus the correspondence principle unifies
the known results on black hole entropies and enables us to understand
many new cases.

Section 2 treats black holes without Ramond-Ramond charges, starting with
Schwarzschild in any dimension, and then including electric
Neveu-Schwarz charges.\footnote
{Black holes with magnetic Neveu-Schwarz charge are the one example in which
the horizon 
does not become smaller than a string at weak coupling.  An approach for
understanding these entropies has been discussed in~\cite{larwil,cvts3}.}
Section 3 treats black
$p$-branes with a single Ramond-Ramond charge.
The near extremal entropy of these solutions have been
discussed previously \cite{klts} where it was argued that in most
cases it could not be understood simply in terms of the known light degrees of
freedom on the brane. We will see that when the correspondence principle 
is applied, the near extremal entropy in all cases agrees with the D-brane
counting (up to factors of order unity)\footnote{For other discussions of 
the near extremal entropy, see \cite{lmpr, mal5, klts2}.}.  We will also
discuss two ways of compactifying  the black $p$-brane spacetime
and show that their entropies are  reproduced by 
two different configurations of
D-branes. Section 4 presents a discussion of the relation to other work
and some concluding remarks. 

\sect{Neutral and NS-NS Black Holes}

\subsection{Schwarzschild Black Holes}

We start with the familiar four dimensional Schwarzschild black hole
\eq \e{schw}
ds^2 =-\left( 1 - {r_0\over r}\right) dt^2 + \left( 1 - {r_0\over r}\right)^{-1}
dr^2 + r^2 d\Omega\ .
\eqe
The mass of the black hole is $M_{bh} = r_0/2G$.
We want to equate this with the mass of a string state at excitation level
$N$, which is $M_s^2 \sim N/\apm$ at zero string coupling $g$. In four
dimensions Newton's constant is related to the string coupling $g$ and
$\apm$ by
$G \sim g^2 \apm$.  So it is clear that the mass of the black hole
cannot equal the string mass for all values of $g$. If we want to
equate them, we have to decide at what value of the string coupling
they should be equal.
Clearly, the natural choice is to let $g$ be the value at which the
string forms a black hole, which, by our correspondence principle is
when the horizon is of order the string scale.
Setting the masses equal when $r_0^2 \sim \apm$ yields
\eq
M_{bh}^2 \sim {\apm \over G^2} \sim {N\over \apm} \e{match}
\eqe
The black hole entropy is then
\eq
S_{bh} \sim {r_0^2 \over G}\sim {\apm \over G} \sim \sqrt N
\eqe
So the Bekenstein-Hawking entropy is comparable to the string
entropy~\cite{suss1}.  They have the same dependence on the mass and only
differ by a factor of order unity which depends on exactly when the
string state forms a black hole.

Put differently, consider following a particular state as the coupling is
varied, which means holding the entropy fixed.  The success of the
above matching means that the mass changes only by a factor of order
one during the transition from the black hole description to the string
description.  There are various large and small dimensionless
numbers in the problem.
One is the excitation level $N$.  Another is the string coupling $g$ at
the transition; from the matching condition~\ee{match} and $G \sim g^2
\apm$, one finds $g \sim N^{-1/4}$.  It could have turned out
that the mass changes
during the transition by a factor which is parametrically large, such as
a power of $N$, but this is not the case here or in any of the later
examples.

Since the string forms a black hole at the string scale, which is not
large compared to the compactification scale, it is important to see
whether  this agreement continues to hold for black holes in higher
dimensions. The Schwarzschild metric in $d$ spatial dimensions is similar
to~\ee{schw} except that $r_0/r$ is replaced by $(r_0/r)^{d-2}$. The mass
is now
$M_{bh} \sim r_0^{d-2}/G$. We again equate this with the string mass when
the black hole is of order the string size
\eq
M_{bh}^2 \sim {(\apm)^{d-2} \over G^2} \sim {N\over \apm}
\eqe
The black hole entropy is thus
\eq
S_{bh} \sim {r_0^{d-1} \over G}\sim {(\apm)^{(d-1)/2} \over G} \sim \sqrt N
\eqe
So once again the black hole entropy is comparable to the string entropy.

One might have been concerned that the typical string state is much
larger than the string scale and so does not sensibly match onto the
black hole~\cite{suss1}.  This is a somewhat involved question, about
which we will have more to say in a future paper \cite{horpol}, but in
fact it is not really relevant here.  A highly excited string is like a
random walk, with an entropy proportional to its length.  Even if we
restrict attention to highly excited string states that are small,
constrained to lie in just a few string volumes, the entropy is still
proportional to the length, just with a numerically smaller coefficient
to which we are not sensitive.  In fact there is an offsetting effect due
to the gravitational self-interaction~\cite{horpol}.

\subsection{Charged Black Holes}

We now consider charges that can be carried by fundamental
strings, i.e. electric Neveu-Schwarz charges associated with momentum
and winding modes.  For a black hole with
these charges, the dilaton is not constant, so the string metric 
differs from the Einstein metric. The correspondence between strings
and black holes occurs 
when the curvature of the string metric at the horizon
is of order the string scale. 
We will see that this implies that the 
size of the horizon in the string metric is of order the string scale.
We will equate the mass of the black hole to the mass of
the string at this point and show that the black hole entropy (which is
proportional to the horizon area in the Einstein metric) is then
comparable to the usual string entropy.

For a string propagating on a circle with radius $R$, the left and right moving
momenta are defined to be 
\eq
p_L = {n\over R} - {mR\over \apm}, \qquad \qquad p_R = {n\over R} +
{mR\over \apm}
\eqe
where $n,m$ are the integer momentum and winding numbers. The string 
entropy is
\eq
S_s \sim \sqrt{N_L} + \sqrt{N_R} 
\eqe
where
\eq
{N_L\over \apm} \sim M_s^2 - p_L^2, \qquad \qquad
{N_R\over \apm} \sim M_s^2 - p_R^2
\eqe

To obtain the four dimensional
black hole solution with these charges we start with
the five dimensional black string solution \cite{host, hhs}
(in the string metric)
\eq\e{bkst}
ds^2 = F\left[ -\left( 1-{r_0\over r}\right) d t^2 + d z^2\right]
+ \left( 1-{r_0\over r}\right)^{-1} dr^2 + r^2 d\Omega
\eqe
where
\eq
F^{-1}= 1+ {r_0 \sinh^2 \gamma_1\over r} 
\eqe
and the dilaton is $e^{2\phi_5} = F$.
One can now boost along the $z$ 
direction (to add momentum) and reduce to four dimensions to obtain
\eq\e{dilbh}
ds^2 = -\Delta^{-1}\left( 1-{r_0\over r}\right) dt^2 + 
 \left( 1-{r_0\over r}\right)^{-1} dr^2  + r^2 d\Omega
 \eqe
$$ \Delta = \left(1+{r_0 \sinh^2 \gamma_1\over r}\right)
             \left(1+{r_0 \sinh^2 \gamma_p\over r}\right)\ . $$
The four dimensional dilaton (which differs from $\phi_5$
by a factor of the length of the fifth dimension) is
$e^{-4\phi} = \Delta$.
The horizon is at $r=r_0$,
and the ADM mass is
\eq
M_{bh}= {r_0\over 8G} (2+\cosh 2\gamma_1 + \cosh 2\gamma_p) 
\eqe
where $G$ is the four dimensional Newton's constant. The integer normalized
charges corresponding to the momentum and winding numbers are\footnote{We
follow the conventions of \cite{mal, hlm}. The left and right boost
parameters of \cite{sen} are related to $\gamma_1, \ \gamma_p$ by
$\alpha = \gamma_p - \gamma_1$ and $\beta = \gamma_p + \gamma_1$.}
\eq
n = {r_0 R \over 8G} \sinh{2\gamma_p}, \qquad \qquad
m= {r_0 \apm \over 8G R} \sinh{2\gamma_1}
\eqe
The left and right moving momenta are thus
\eqa
p_L &=& {r_0\over 8G} (\sinh 2\gamma_p -\sinh 2\gamma_1) \cr
p_R &=& {r_0\over 8G} (\sinh 2\gamma_p +\sinh 2\gamma_1)
\eqae
The horizon area in the Einstein metric, $ds_E^2 = e^{-2\phi} ds^2$, is
\eq\e{areac}
A = 4\pi r_0^2\cosh\gamma_1 \cosh \gamma_p.
\eqe

The curvature of the full ten dimensional string metric (which
is the product of~\ee{bkst} with a five torus) has two independent
components near the horizon. One is proportional to $1/r_0^2$ and the other is 
proportional to $(\tanh^2 \gamma_1)/r_0^2$. So the first is always larger,
and the curvature is of order the string scale when $r_0^2 \sim \apm$.
Setting the mass and charges of the black hole equal to those  of the
string at this scale yields
\eqa\e{ndef}
{N_L\over \apm} \sim  M_{bh}^2 - p_L^2 \sim {\apm \over G^2}
[3+2(\cosh 2\gamma_1 +\cosh 2 \gamma_p) + \cosh 2(\gamma_1 + \gamma_p)] \cr
{N_R\over \apm} \sim  M_{bh}^2 - p_R^2 \sim {\apm \over  G^2}
[3+2(\cosh 2\gamma_1 +\cosh 2 \gamma_p) + \cosh 2(\gamma_1 - \gamma_p)] \cr
\eqae
The largest term or terms in $N_L$ or $N_R$ is always of order
$\apm^2 G^{-2} \cosh 2(|\gamma_1| + |\gamma_p|)$, which for all $\gamma_1,
\gamma_p$ is
the same as $\apm^2 G^{-2} \cosh^2 \gamma_1 \cosh^2 \gamma_p$ up to a
factor of order one.  The string entropy is then
\eq
S_{s} \sim \sqrt {N_L} + \sqrt {N_R} \sim {\apm\over G} 
\cosh\gamma_1 \cosh \gamma_p  
\eqe
This is the same as the black hole entropy $S_{bh} \sim A/G$, where the area
is given in eq.~\ee{areac}, at the point $r_0^2 \sim \apm$ where the matching
is done.  Thus we find that the black hole entropy always agrees with the
string entropy up to factors of order unity.  In particular, it has the same
dependence on the mass and charge. We have also checked that this agreement
extends to charged black holes in higher dimensions.

It should be noted that even after the transition to the weakly coupled
regime, the gravitational dressing remains large.  The stringy behavior at 
$r \sim r_0$ smears out the zero in $(1-r_0 / r)$ and so this does not
cause a
large correction (essentially, this is part of the correspondence
principle). But the factor $\Delta$ differs significantly from unity over a
much greater scale when $\gamma_1$ or $\gamma_p$ is large, and its value near
the horizon is of order $\cosh^2 \gamma_1 \cosh^2 \gamma_p$.  
We must use the corrected local metric in calculating the string entropy.
This does
not, however, affect the result.
Consider first the case $\gamma_p = 0$, so there is only winding charge;
the dressing~\ee{bkst} is then a uniform rescaling of the $zt$ plane.
Near the extremal limit, the mass relation becomes  
\eq
M_{bh} - \frac{mR}{\apm} \sim \frac{N }{ m R}.
\eqe
The left-hand side is the free energy, the excess energy above the rest
mass of the winding strings.  Its value near the string is greater than its
asymptotic value by the redshift  $\cosh \gamma_1$.  But also the
radius $R$ is contracted by the same factor, so the value of $N$ and hence
the entropy are the same as would follow from the asymptotic values.
For the second charge, the compact momentum, we do not need a detailed
analysis: it is simply a result of applying a boost  to both the black
hole and string configurations.

The redshift does have one notable effect. 
The asymptotic temperature at the
matching point is 
$T \sim (\apm^{1/2}\cosh\gamma_1\cosh\gamma_p)^{-1}$.  With the redshift
included, the local temperature at the
string is $T \sim (\apm)^{-1/2}$, the string scale, just as in the
Schwarzschild case~\cite{suss1}.

String states with $N_R=0$ are supersymmetric. On the black hole side
this corresponds to an extremal limit $r_0/G \rightarrow 0,\ \gamma_p, \gamma_1
\rightarrow
\infty$ keeping $(r_0/G)\sinh 2 \gamma_1 $ and $\gamma_1 -\gamma_p$ fixed.
In Planck units (constant $G$), the horizon area~\ee{areac} vanishes
in this limit,  which led Sen \cite{sen2} to compare the number of
string states with the area of a 
`stretched horizon' where the curvature of the extremal solution
was of order the string scale.
Since we start with nonextremal black holes and set the size of the
horizon in the string metric to be of order the string scale,\footnote{Since
$r_0 \sim (\apm)^{1/2}$, the string coupling $g$ must become large as $r_0/G$
vanishes.}
the entropy
remains finite in the limit and agrees with the string expression.

\sect{Black $p$-Branes}

\subsection{Ten Dimensions}

In this section we consider black $p$-branes with a single Ramond-Ramond
charge. The string metric is given by \cite{host}
\eq\e{blkp}
ds^2 = f^{-1/2}
\left[ -\left( 1-{r_0^n \over r^n}\right) d t^2 + d y^i dy_i\right]
+ f^{1/2}
\left[\left( 1-{r^n_0\over r^n}\right)^{-1} dr^2 + r^2 d\Omega_{n+1}\right]\ .
\eqe
where
\eq\e{fdef}
f = 1 + {r_0^n \sinh^2 \alpha \over r^n}\ .
\eqe
The $y_i$ are $p=7-n$ spatial coordinates along the brane
which we assume are compactified on a large torus of volume $V$.
The dilaton is $e^{2\phi} = f^{(n-4)/2}$.
The energy, RR charge, and entropy of the $p$-brane are
\eqa
E &\sim & {r_0^n V \over g^2 \apm^{4}} \left({n+2\over n} + \cosh 2\alpha
\right)\nonumber\\
Q &\sim & {r_0^n \over g \apm^{n/2}} \sinh 2\alpha \e{eqs}\\
S_{bh} &\sim& {r_0^{n+1} V \over g^2 \apm^{4}} \cosh \alpha \nonumber
\eqae
We have dropped overall constants of order unity since they will not
be needed for testing
the correspondence principle. However, it should be noted that
the constants in front of $E$ and $Q$ are the same, so that
in the extremal limit ($r_0 \rightarrow 0$, $ \alpha \rightarrow \infty$,
with $Q$ fixed) we have $E = QV/g \apm^{(8-n)/2}$.

In the limit of weak string coupling, this extremal limit corresponds to 
$Q$ Dirichlet $p$-branes \cite{pol}. The nonextremal solution should 
correspond to 
an excited state of these D-branes and strings. To determine when
this weak coupling description is applicable, we consider
the curvature of~\ee{blkp} at the
horizon $r = r_0$. The largest contribution
comes from the angular part of the metric and is of order 
$(r_0^{2} \cosh\alpha)^{-1}$. By the correspondence principle, the matching
between the black $p$-brane and  the strings and D-branes occurs when
this is of order $1/\apm$ or
\eq
r_0 = \frac{\apm^{1/2}}{(\cosh\alpha)^{1/2}}\ .\e{bpmatch}
\eqe
That is, for given $Q$ and $S$, eq.~\ee{bpmatch} determines the value of
$g$ at which the description changes.
At this point we wish to compare
the Bekenstein-Hawking entropy to that of an assembly of strings
and D-branes with the same
charge and mass. Note that since
$e^{\phi} = (\cosh\alpha)^{(n-4)/2}$
on the horizon,  eq.~\ee{bpmatch} implies that $g e^{\phi} < 1/Q$. Thus
the local string coupling remains small.

There are two qualitatively different kinds of excited states of D-branes. 
The first consists of adding
a small number of long strings.\footnote
{The black $p$-brane should match onto long strings lying in approximately the
volume of the $p$-brane, but this constraint will only affect the
coefficient in the entropy.  Also, long open strings ending on the D-branes
are more
numerous (and hence more likely) than long closed strings,
but the effect on the entropy is
of subleading order.}
In this
state the entropy is $S_1 \sim \apm^{1/2}\Delta E$,
 where 
\eq
\Delta E = E - {QV\over g \apm^{(p+1)/2}} 
\eqe
is the excess energy above the D-brane rest mass.  The second class of
states consists of exciting a large number of massless
open strings on the D-branes.
There are $Q^2$ species of open string, as will be explained below, so
the excess energy and the entropy (dropping numerical constants) are
\eqa\e{gasent}
\Delta E &\sim& Q^2 T^{p+1} V \nonumber\\
S_2 &\sim& Q^2 T^{p} V 
\eqae
which implies
$S_2 \propto (\Delta E)^{p/(p+1)}$.
Not surprisingly, for large excess energy, long string states are more
numerous, while the string gas has higher entropy when  $\Delta E$ is small.
The transition occurs when $\Delta E \sim Q^2 V/ \apm^{(p+1)/2}$, i.e.,
when $T$ is of order one in string units.

We now wish to compare these weak coupling entropies with $S_{bh}$. 
The 
procedure is to match the energy and the charge of the strings and D-branes
with that of the black $p$-brane
when~\ee{bpmatch} is satisfied.  At this point, 
\eqa\e{eqmat}
\Delta E &\sim& {V \over g^2 \apm^{(p+1)/2}} (\cosh\alpha)^{-n/2}
\nonumber\\
Q &\sim& {1 \over g} \sinh 2\alpha \ (\cosh\alpha)^{-n/2}
\eqae
and so
\eqa
S_1 &\sim& {V \over g^2 \apm^{p/2}} (\cosh\alpha)^{(p-7)/2}
\nonumber\\
S_2 &\sim& {V \over g^2 \apm^{p/2}} 
(\tanh \alpha)^{2/(p+1)} (\cosh\alpha)^{(p-6)/2}\ .
\eqae
Large  $\alpha$ corresponds to near extremal configurations (with $S_2 >S_1$),
while small $\alpha$ corresponds to configurations far from extremality
(with $S_1 > S_2$). In either case, the larger of the two entropies
agrees with the black hole entropy, which from~\ee{eqs} and \ee{bpmatch}
is
\eq
S_{bh} \sim {V \over g^2 \apm^{p/2}} (\cosh\alpha)^{(p-6)/2}\ .
\eqe
Thus the correspondence principle correctly reproduces the entropy of 
all black $p$-branes
with a RR charge. 

As in the case of near extreme black holes with NS charges,
there is a large red-shift near the D-branes, but it does not affect the result.
 From the metric~\ee{blkp} $f$ is a
uniform rescaling of the D-brane world-volume, to which the massless string
gas is insensitive.  In other words, in the ideal gas relations~\ee{gasent}
we need to include a redshift factor $\gamma = (\cosh\alpha)^{1/2}$ in $\Delta
E$ and
$T$ and $\gamma^{-p}$ in the volume, so these equations continue to hold.
The redshift again raises the  
asymptotic temperature at the matching point,
$T \sim (\apm\cosh\alpha)^{-1/2}$, to the string scale $(\apm)^{-1/2}$.

Now we must justify the assumption of $Q^2$ degrees of freedom in the
gas regime.  The
massless fields on the D-branes are the non-Abelian gauge fields and
collective coordinates \cite{tasi}.  These are $Q^2$ in number, but the
collective coordinates have a potential proportional to Tr$([X^i, X^j]^2)$
so that the moduli space is of dimension $Q$; similarly the gauge fields
have a self-interaction. These interaction terms will restrict the number
of effective degrees of freedom only if they are large compared to the
kinetic term, which turns out not to be the case. 
Treating the interaction as a perturbation on a
free gas of $Q^2$ species, we can estimate its ratio to the kinetic term
as follows.  By the usual large-$N$ counting, the ratio will have a
factor of $ge^{\phi}Q$.  The quartic coupling $g$ in $p+1$
spacetime dimensions has units of (mass)$^{3-p}$, so
the dimensionless expansion parameter is $g e^{\phi} Q T^{p-3}$.  We have seen
that at the horizon $T \sim (\apm)^{-1/2}$ and
$g e^{\phi} Q \sim (\apm)^{(p-3)/2}$, so the expansion parameter is of order
one.\footnote
{Curiously the position-dependence of $\phi$ and of the effective $T$ cancel,
so one comes to the same conclusion by erroneously using the asymptotic
values.} That is, the two energies are of the same order.  The potential is
positive so we have underestimated the energy of the gas, but only by a
numerical factor.  This is within the accuracy
of the correspondence principle.\footnote{The gas picture doesn't apply for
$p=0$, but the result is the same: the potential determines the magnitude of
the fluctuations of the
$Q^2$ degrees of freedom.}

The correspondence principle is thus confirmed for the large class of 
solutions \ee{blkp},  but we should
note that it is working even better than one might expect in some
cases. Specifically, for $n > 4$ ($p = 0,1,2$) and $\alpha$ large, spheres
outside the horizon
$r = r_0$ grow {\it smaller} as $r$ increases, reaching a minimum size
at $r \sim r_0
(\sinh\alpha)^{2/n}$.  Correspondingly, the maximum curvature outside the
horizon occurs at a finite distance away from the horizon.
Thus as one decreases the string coupling, the
$\apm$ expansion first breaks down away from the horizon, and
there is a range of couplings where the horizon and asymptotic
region are both
described by low energy gravity but an intermediate region has
string corrections.
In applying the correspondence principle we have implicitly
assumed that the expressions for the black $p$-brane~\ee{eqs} remain valid until
the curvature at the horizon itself is of order the string scale.
It is not clear why this is justified.  Incidentally, 
if one attempts to match the black hole to the D-branes
when the curvature away from the horizon first becomes of order the string
scale, one finds that the entropies do not agree---the D-brane entropy is
too small.

\subsection{Compactification}

We now consider compactification of the black $p$-branes below ten dimensions.
We begin with the simplest case of zero charge ($\alpha=0$),
when the solutions are
just the product of a torus $T^p$ and the $10-p$ dimensional
Schwarzschild metric.
This is one form of compactification, and was considered in section 2. 
However there is another possibility: One can consider a $p$-dimensional 
array of ten
dimensional Schwarzschild black holes. An  array of finite size would not
be static, but an infinite array does lead to a static solution 
\cite{myers}. Identifying after one period, the array has the
same mass and entropy as a single ten dimensional black hole
$M\sim r_0^7/g^2 \apm^4$
and $S\sim r_0 M$, while the product solution has $M'\sim\rho_0^n V/g^2 \apm^4$,
and $S'\sim  \rho_0 M'$ where $\rho_0$ is the Schwarzschild radius of the lower
dimensional black hole, $V$ is the volume of the internal space, and
$n=7-p$. Setting the
masses equal $r_0^7 \sim \rho_0^n V$, yields
\eq
{S\over S'} \sim {r_0\over \rho_0} \sim \left({V\over r_0^p}\right)^{1/n}
\eqe
So the array has greater entropy as long as $V>r_0^p$, below which point
the images start to merge. This suggests that the product solution
is unstable in this regime,
a fact which has been confirmed by studying the linearized
perturbations \cite{grla}.

Both the entropy of the product solution and the array can be understood
by our correspondence principle, since we saw in section two that it
reproduces the entropy of Schwarzschild black holes in any dimension. 
The difference between the two cases is the following: 
As we decrease the string coupling, the
mass of the black hole in string units increases. 
For the product solution, the curvature
at the horizon reaches the string scale at a larger value of the coupling
than for the array. Hence the energy of the resulting string
state and associated entropy is smaller.

Now we consider the near extremal solutions $\alpha \gg 1$.
(Solutions far from extremality
are qualitatively similar to the case of zero charge.)
We first consider
a periodic coordinate transverse to
the $p$-brane, say $x^9 \sim x^9 + 2\pi R$.  Again there are two kinds of
black hole, the array and the translationally invariant solution.
For the extremal black
$p$-brane, the array is given by replacing
$f$ in \ee{fdef} with
\eq
f_1 = 1 + r_0^n\sinh^2 \alpha_1 \sum_{k = -\infty}^\infty
{1\over |x - x_k|^n}\ , \e{array}
\eqe
where $x^9_k = 2\pi R k$ are the image positions, and taking $r_0 \to 0$
with  $r_0^n \sinh^2 \alpha$ fixed.  The array of
nonextremal solutions, $r_0 >0$, is more complicated \cite{lpx}, 
but for $r_0 \ll R$
it is  easy to construct an approximate solution. The metric for a single
nonextreme
$p$-brane is indistinguishable from the extreme solution when $r \gg r_0$.
So one can approximate the nonextreme array by keeping
$f_1$ as above and inserting factors of
$(1-{r_0^n / r^n})$ just as in the ten-dimensional solution~\ee{blkp}
(where $r$ is a radius from each $p$-brane).

The homogeneous solution,
which is translationally invariant in the $x^9$ direction, is
\eqa\e{transol}
ds^2 &=& f_2^{-1/2}
\left[ -\left( 1-{\rho_0^{n-1} \over \rho^{n-1}}\right)dt^2 + d y^i dy_i\right]
\\
&&\qquad\qquad+ f_2^{1/2}
\left[\left( 1-{\rho_0^{n-1} \over \rho^{n-1}}\right)^{-1} d\rho^2 +
\rho^2 d\Omega_{n} + dx_9^2 \right]\ ,\nonumber
\eqae
where
\eq
f_2 = 1 + {\rho_0^{n-1} \sinh^2 \alpha_2 \over \rho^{n-1}}
\eqe
The array solution~\ee{array} has the same energy, charge and entropy as the
ten-dimensional solution, eq.~\ee{eqs}, while for the homogeneous
solution~\ee{transol} these are
\eqa\e{invsol}
E' &\sim& {\rho_0^{n-1} R V \over g^2 \apm^{4}} \left({n+1\over n-1}
+ \cosh 2\alpha_2\right)
\nonumber\\ 
Q' &\sim& {\rho_0^{n-1} R \over g \apm^{n/2}} \sinh 2\alpha_2
\e{eqt}\\ 
S'_{bh} &\sim& {\rho_0^{n} R V \over g^2 \apm^{4}} \cosh \alpha_2
\nonumber
\eqae
 For large $R$, one would expect the
compactification to have little effect and so the array
solution~\ee{array} would appear to be more physical.  As before,
we can determine which solution is stable by seeing which has
more entropy for given mass and charge.
 For equal masses and
charges, eqs.~\ee{eqs} and~\ee{eqt} imply that
$\alpha_1 \sim \alpha_2$ and $\rho_0^{n-1} R \sim r_0^n$.  It follows that 
\eq
\frac{S_{bh}}{S'_{bh}} \sim \frac{r_0}{\rho_0} \sim 
\left(R\over r_0\right)^{1/(n-1)}
\ .
\eqe
Once again,
the array has greater entropy as long as $R > r_0$, below which point
the array solution approaches the translationally
invariant one. Notice that in the near extremal limit, $r_0$ is very
small, so the homogeneous solution is almost always unstable.

Let us now consider the weak coupling description in terms of D-branes.
The array corresponds to
$Q$ coincident D-branes.  The translationally invariant solution
corresponds to $Q$ D-branes evenly distributed in $x^9$.
To count the number of excited states, it is convenient to apply
$T$-duality.  The D $p$-branes then become $(p+1)$-branes, extended in the
$x^9$ direction.  The $x^9$ coordinate is $T$-dual to the D-brane Wilson
line \cite{tasi}, so for the coincident D-branes the Wilson line is the
identity while for the distributed D-branes its eigenvalues are uniformly
distributed.  In the latter case, one can go to a basis in which the
Wilson line is the shift matrix,
\eq
W_{ij} = \delta_{i,j+1}, \quad (i  \equiv i + Q).
\eqe
With this Wilson line the D-brane fields $\phi_{i,j}$ (both the collective
coordinates and the gauge fields) are periodically identified with
$\phi_{i+1,j+1}$ and one essentially has
$Q$ species (distinguished by $|i-j|$) on a D-brane of length $2\pi Q R' = 2\pi
Q\apm/R$.  We refer to this as the wrapped system since it describes
one D-brane wrapped $Q$ times around the circle \cite{dama}.  For $W = 1$, the
unwrapped system, there are
$Q^2$ species on a D-brane of length
$2\pi R'$.

For $TR' > 1$, meaning large $R'$ or large energy density, these two
systems behave essentially the same, with
\eqa
\Delta E &\sim& Q^2 V R' T^{p+2} \nonumber\\
S &\sim& Q^2 V R' T^{p+1}.
\eqae
When $TR' < 1$, 
the modes of the unwrapped system can no longer propagate
in the compact direction and the system behaves like a $p$-dimensional
gas, while the wrapped system continues to behave like a $(p+1)$-dimensional
gas. We now show that when the correspondence principle is applied,
these two systems reproduce the entropy of the array and homogeneous
black $p$-branes respectively. The entropy of the array is the same as a single
black $p$-brane in ten dimensions, since the compactification has little effect.
We have already seen that its near extremal entropy is reproduced by a
gas of $Q^2$ degrees of freedom  in $p$ dimensions which is just the
unwrapped system. If we apply T-duality to the homogeneous solution,
the effect on the metric \ee{transol} is $g_{99} \to 1/g_{99}$ and
we obtain the solution for a black $(p+1)$-brane. The energy and entropy
are duality invariant. Thus the near extremal entropy
is reproduced by a gas of $Q^2$ degrees of freedom in $(p+1)$-dimensions
which is the wrapped D-brane. Notice that the wrapped D-brane
has {\it less} entropy than the unwrapped one.

What is the role of the transition point $TR' \sim 1$ in the black $p$-brane
picture? As we have seen, the asymptotic
temperature satisfies $T \sim (\apm\cosh \alpha)^{-1/2}$, so the radius
above which the wrapped and unwrapped systems become indistinguishable
is $R' \sim (\apm\cosh \alpha)^{1/2}$.  The $T$-dual radius is $R = 
(\apm/\cosh \alpha)^{1/2} = r_0$, which is just the radius below which the
array overlaps into a translationally invariant system.  Thus we see a
detailed correspondence between wrapping/unwrapping for D-branes and the
two kinds of compactified black holes.

Now let us consider compactification of one of the directions parallel to
the $p$-brane, i.e. we suppose one direction is much smaller than the rest. 
To make use of the previous discussion (to which it is
$T$-dual), we start with a black
$(p+1)$-brane, compactify with periodicity $2\pi R'$, and denote the
volume of the $(p+1)$-brane by $2\pi R' V$. 
We saw  in section 3.1  that the near extremal entropy  is reproduced
by a $(p+1)$-dimensional gas with $Q^2$ species.
For $R'T >1$ this can be represented in weak coupling either by
wrapped or unwrapped D-branes. However, for $R'T<1$, only the wrapped
system yields the black $(p+1)$-brane entropy; the unwrapped branes
have higher entropy.
It may seem  puzzling that the black $(p+1)$-brane
ceases to be the lowest-entropy configuration at a rather large radius,
$R'  \sim (\apm\cosh \alpha)^{1/2}$. Moreover, the
higher-entropy configuration is hard to describe.  It is $T$-dual to the
array.  But the array is not invariant under $x^9$ translation---that is,
the background has modes of nonzero $p_9$---so the dual background must
have fields associated with nonzero winding number. 
We can see a signature of this in
the black $p$-brane metric~\ee{blkp}.  The metric along the D-brane, at
the horizon, is $dy^2 (\cosh \alpha)^{-1/2}$, so that when $R' \sim
(\apm\cosh \alpha)^{1/2} $ the size of the compactified direction {\it at
the horizon} is only $\apm^{1/2}$ and so it is possible for stringy
effects to arise.

It is interesting to note if we continue to reduce the radius until
$QR'T <1$, the entropy of the wrapped D-brane changes from a $p+1$
dimensional gas to a $p$ dimensional gas with $Q$ degrees of freedom.
This is to be expected since the T-dual configuration now consists of
$Q$ widely separated D-branes. The strong coupling limit of this  would
be $Q$ near extremal black $p$-branes each with unit charge. Thus
when $QR'T <1$, the entropy of the homogeneous $p$-brane is not only smaller
than the array of charge $Q$ $p$-branes, but also smaller than an array
of charge one $p$-branes with spacing $1/Q$ of the previous period.

To summarize, we have seen that the correspondence principle works 
in great detail: we have considered two different kinds of compactified black
$p$-branes and two configurations of D-branes,
and the entropies match in detail both
for the higher-entropy and the lower-entropy system.

\subsection{Black Holes with Two or More RR Charges}

Recently, a precise agreement (including the numerical coefficient)
has been found \cite{stva,homa} between the entropy of certain
extreme and near extreme black holes and states of D-branes. These
cases differ from the ones we have discussed in that there are 
at least two Ramond-Ramond charges.  An example is the
five-dimensional black hole carrying one-brane charge, five-brane charge, and
compact momentum.  
Applying the correspondence principle to this black hole, there are eight
cases to consider according to which of the three charges are large.  These
separate into three categories.  When neither of the Ramond charges is large
the results of section~2 apply: the typical weak-coupling state is a long
string, whose entropy matches that of the black hole.  When one of the Ramond
charges is large, the discussion from earlier in this section applies and the
black hole entropy matches that of an interacting string gas with $Q_1^2$ or
$Q_5^2$ degrees of freedom.  When both Ramond charges are large, 
the entropy is reproduced precisely by a gas of $Q_1 Q_5$ moduli
\cite{stva,homa,nonex}.  This is now a free gas, in that the moduli have no 
potential.\footnote{There is a puzzle here, in that the local temperature
at the transition is of order the string scale and so large enough that
there will be some excitation of states other than the moduli.  This does
not affect the qualitative agreement required by the correspondence
principle, but makes the precise agreement for near extremal entropies
\cite{nonex} somewhat puzzling.}

Note that in most of the cases we have discussed, if one sets the energy
and charge of the weak coupling state equal to that of the black hole
at an arbitrary value of the Schwarzschild radius $r_0$,
the two entropies have different
dependence on $r_0$. 
For example, in the Schwarzschild black hole $S_{bh} \propto
r_0^2$ while $S_{str} \propto r_0$ at fixed $g$.  The matching of entropies
then depends on a special value of $r_0$ given by the correspondence
principle.   For the
cases where exact calculations of the entropy have been done, the $r_0$
dependence is the same on both sides and so the matching scale drops out.

\sect{Discussion}

We have proposed a correspondence principle which connects black holes
to weakly coupled strings and D-branes, and shown that it leads to an  
agreement between the entropy of these  two systems. Although
we have not been able to compare the precise coefficients in the entropy
formulas, they have the same dependence on the mass and charge in a 
wide variety of different contexts. 
This strengthens the idea that a black hole is an ordinary quantum
mechanical system, and that string theory is a viable theory of quantum
gravity.  In all examples considered here, string theory provides the
correct number of degrees of freedom to account for the black hole entropy.

We have seen that the typical string state depends in an essential way on the
quantum numbers.  With no large RR charges it is a single long string, with
one it is an interacting string gas on D-branes, and for some examples 
with two or more RR charges it is a
free string gas, a gas of moduli.  The correspondence principle
thus unifies various results in the literature.

The success of the correspondence principle does not mean that
gravitational effects remain small whenever the string and D-brane
picture is valid. For near extremal black holes, we have seen that
the metric deviates from flat space over a region much larger than
the horizon size. Thus there is a large gravitational dressing  
after the transition. This affects both the local energy and the size
of the internal space, but suprisingly, the entropy is unaffected.
In all cases, the local Hawking temperature at the matching point is 
of order the string scale.

One can trivially extend this agreement in certain ways, e.g. by
adding momentum to the black $p$-branes discussed in section 3. The only
change in both the black hole and weak coupling descriptions is to apply
a boost in some direction. Other extensions are probably possible, but
require further investigation. For example consider magnetic NS charge.
It is clear that when the charge is small, it has
little effect on the entropy and the matching can be done as in section 2.
In other cases, e.g. the near extremal black five-brane in IIB string theory
with large charge, the correspondence principle cannot be applied directly since
the horizon size never becomes small in string units. However in this
case, 
the string coupling becomes large near the horizon. It thus 
seems appropriate to count states  by going to the weakly coupled 
S-dual description.
This is five-brane with RR charge, for which the correspondence
principle can be applied and yields the correct entropy. 

It appears difficult to extend our analysis to try to compare the precise
coefficients in the formulas for the entropy. 
This would require a better
understanding of the string state when it is of order the string scale.

Near extremal black $p$-branes with one RR charge
have been discussed recently in the literature. It is perhaps useful
if we comment on the relation between our discussion in section 3 and
some of this previous work.
It has been noted that the black hole and D-brane entropies have different
temperature (or equivalently $r_0$) dependence in general~\cite{klts}.
This is in accord with our point of view, since we expect them to
match only at one point.\footnote{In
ref.~\cite{lmpr} it is shown that higher dimension
operators in the action can in a rather general way correct the
temperature dependence of the black hole entropy to that of the
D-brane gas.  Possibly there is some relation to our work, though the
physical assumptions are rather different.}

The self-dual case $p=3$ is particularly
interesting.  If we consider the $r_0$ dependence of the entropy at
fixed $g$ and $\alpha$, the black hole entropy is proportional to $r_0^{8-p}$
while the string gas entropy is proportional to $r_0^{(7-p)(p+2)/(p+1)}$.
Precisely for $p=3$ these are the same, so the matching scale drops out and
one might hope to relate the entropies precisely .  However, this is
the  case where the D-brane entropy seems to exceed that of the black hole by a
factor of 4/3 \cite{klpe}.  We now have some understanding of the origin of
this factor.  The D-brane calculations were done with a gas of $Q^2$ species
treated as free.  We have seen, however, that the interactions are of order~1;
neglecting them underestimates the energy of each state
by a factor of order 2 and so
overestimates the entropy (at fixed energy) by a similar factor.  The
interactions are quite complicated, however, and we do not see a way to
obtain the precise factor.

In the case $p=5$, the black $p$-brane entropy has been shown to agree
precisely with that of a gas of non-critical closed strings living on the
five-brane \cite{mal5}.  We, on the other hand, have shown approximate
agreement with  the entropy of ordinary D-branes and open strings.  Moreover,
in the latter case the open string interactions are of order one and the
gravitational dressing is large, while all such complications are blithely
ignored in ref.~\cite{mal5}.  Is there some duality here?

Susskind has pointed out that our results give an approximate verification
of string duality for non-BPS states.  Many of our examples are related
by duality.  The simplest is just the Schwarzschild black hole, which might
turn into a heterotic string at small $g$ and a type II string at large.  The
agreement of the nonextremal entropies of each string with that of the black
hole implies agreement with each other.  In other words, we can follow a
given state from a long heterotic string, to a black hole, to a long type II
string.  The new ingredient is that the black hole description gives a known
$g$-dependence of the mass in the intermediate regime.

Since the different weakly coupled string theories have different 
degeneracy of states, one might wonder whether they 
could all be consistent  with the Bekenstein-Hawking entropy at a precise
level (when the coefficients are better understood).
There are at
least two ways in which this could occur. The first is if the transition 
from the black hole to the string state takes place at slightly different
values of the curvature in the different string theories. The second is
that, as we have remarked several times, the black hole state may turn
into only a subset of the available string states. This subset is 
large enough so that its entropy differs from the usual string entropy
only by an overall coefficient.
In this sense, there may not be a precise one-to-one correspondence between
string states and black holes, and may explain the factor of two
discrepancy in \cite{suss2}.

\subsection*{Acknowledgments} 
We wish to thank J. Rahmfeld and L. Susskind for discussions.
This work was supported in part by NSF grants PHY91-16964, PHY94-07194, and
PHY95-07065.

\end{document}